\magnification 1200
\baselineskip 16pt

\vskip .8cm
\centerline{\bf PROTECTIVE MEASUREMENTS OF TWO-STATE VECTORS}

\vskip .8cm
\centerline{ \bf Yakir Aharonov$^{a,b}$, and Lev Vaidman$^a$}

\vskip 1cm
\centerline{\it $^a$  School of Physics and Astronomy} 
\centerline{\it Raymond and Beverly Sackler Faculty of Exact Sciences} 
\centerline{\it Tel-Aviv University, Tel-Aviv, 69978 ISRAEL}
 
\vskip .5cm 
 
\centerline{\it $^b$ Physics Department, University of South Carolina}
\centerline{\it Columbia, South Carolina 29208, U.S.A.}

\vskip 1.5cm 

 A recent result about measurability of a quantum state of a
single quantum system is generalized to the case of a single pre- and
post-selected quantum system, described by a two-state vector. The
protection required for such measurement is achieved by coupling the
quantum system to a pre- and post-selected protected device yielding
a nonhermitian effective  Hamiltonian.

 \vskip 2cm

\break

 We present here a point of contact between  two approaches which have
been 
main directions of our research in the recent years. Our numerous
discussion of these subjects with Abner Shimony,  whom we thank  for his crystal-clear thinking,  made it possible for us to
see these issues as they presented here.

Recently it  has been shown that  {\it protective
measurements}$^{1,2}$ can be used for 
``observing''  the quantum state of a single system.  Also, in recent
years an approach has been developed  in which a quantum system is
described, at a given time, by two (instead of one) quantum states: the
usual one evolving toward the future and the second evolving backwards in
time from a future measurement.$^{3-7}$ In this approach, the vector
describing a quantum system at a given time consists of two states.  The
protective measurements$^{1,2}$ are not suitable for observing  two-state
vector. Here we will present a method for measuring two-state vectors of a
single (pre- and post-selected) system.  We shall precede the explanation
of our method by brief reviews of the method of protective measurements of
a single quantum state and of the two-state vector formalism.

The basic protection procedure is introducing a protective potential
such that the quantum state of the system will be a nondegenerate
eigenstate of the Hamiltonian. Let us consider a particle in a
discrete nondegenerate energy eigenstate $\Psi (x)$. The standard von
Neumann procedure for measuring the value of an observable $A$
involves an interaction Hamiltonian,
$$
 H = g(t) P A,\eqno(1)
$$
where  $P$ is the conjugate momentum of pointer variable $Q$, and the 
coupling parameter $g(t)$ is normalized to $\int g(t) dt =1$. The initial state
of the pointer is taken to be a Gaussian centered around zero.
 In standard impulsive measurements, $g(t) \neq 0 $  for only a very
short time interval.  Thus, the interaction term dominates the rest of
the Hamiltonian, and the time evolution
$
e^{-i P A}
$
leads to a correlated state: eigenstates of $A$ with eigenvalues $a_n$
are correlated to measuring device states in which the pointer is
shifted by these values $a_n$. (Here and below we use units such that
$\hbar =1$.)  By contrast, the protective
measurements of interest here utilize the opposite limit of extremely
slow measurement.  We take $g(t) = 1/T$ for most of the time $T$ and
assume that $g(t)$ goes to zero gradually before and after the period
$T$. We choose the initial state of the measuring device  such
that the momentum  $P$  is
bounded.  We also assume that $P$ is a constant of motion not only of the
interaction Hamiltonian (1), but of the whole Hamiltonian. For $g(t)$ smooth enough we obtain an adiabatic process in
which the particle cannot make a transition from one energy eigenstate
to another, and, in the limit $T \rightarrow \infty$, the interaction
Hamiltonian does not change the energy eigenstate.  For any given value of
$P$, the energy of the eigenstate shifts by an infinitesimal amount
given by the first order perturbation theory:
$\delta E = \langle H_{int}  \rangle  = 
\langle
A\rangle P/ T.
$ The corresponding time evolution $ e^{-i P \langle A\rangle} $ shifts the
pointer by the average value $\langle A
\rangle $.  
  By measuring the averages of a sufficiently large number of
variables $A_n$, the full Schr\"odinger wave $\Psi (x)$ can be
reconstructed to any desired precision.
 
Let us turn to the review of the two-state vector formalism
 originated by  Aharonov, Bergmann and Lebowitz$^3$ who considered measurements performed on
a quantum system between two other measurements, results of which were
given. The quantum system  between two
measurements is described by  two states: the usual one, evolving towards the future
from the time of the first measurement, 
and a second state  evolving backwards in time, from the time of the second 
measurement.
 If  a
system  has been prepared at  time $t_1$ in a state $|\Psi_1\rangle$ and  is found
 at  time $t_2$ 
in a state $|\Psi_2\rangle$, then at time $t$, $t_1<t<t_2$, the system is described
by 
 $
\langle \Psi_2 | e^{i\int_{t_2}^{t} H dt}
{\rm ~~~ and~~~}
e^{-i\int_{t_1}^{t} H dt} |\Psi_1\rangle .
$
 For simplicity,  we shall consider the free Hamiltonian to
be zero; then, the system at time $t$ is described by the two states 
 $
\langle \Psi_2 |
$ and
$
|\Psi_1\rangle
$.
In order to obtain such a system, we prepare an ensemble of systems in the
state $ |\Psi_1\rangle$, perform a measurement of the desired variable
using separate measuring devices for each system in the ensemble, and
perform the post-selection measurement. If the outcome of the
post-selection was not the desired result, we discard the system and the
corresponding measuring device. We look only at measuring devices
corresponding to the systems post-selected in the state $\langle \Psi_2 |$.

 The basic concept of the  two-state   approach, the weak value of a physical variable $A$ in the time
interval between pre-selection of the state $| \Psi_1 \rangle$ and
post-selection of the state $ | \Psi_2 \rangle$ is given by$^5$
 $$
 A_w \equiv {{\langle \Psi_2 | A | \Psi_1 \rangle}
\over {\langle \Psi_2 |\Psi_1 \rangle}} ~~~~.
\eqno(2)
$$ 
Weak values emerge from a measuring procedure
with a sufficiently weak interaction.
When the strength of the coupling to the measuring device goes to zero, the
outcomes of the measurement invariably yield the weak value. To be more
precise, a measurement  yields the real part of the weak value. Indeed,
the weak value is,
in general,  a complex number, but its imaginary part will contribute only a
phase to the wave function of the measuring device in the position
representation of the pointer. Therefore, the imaginary part will not affect the probability distribution
of the pointer position, which is what we see in a  usual measurement.
However, the imaginary part of the weak value also has physical meaning. It
expresses itself as a change in the conjugate momentum of  the pointer
variable.

We are familiar with weak measurements performed on a single system. In
fact, the first work on weak measurements$^4$ considered such a
case. There, a single measurement of the spin component of a spin-$N$
system yielded the ``forbidden'' value $\sqrt{2} N$ with the uncertainty $
\sqrt{N}$. This is the weak value of $S_\xi$ for the two-state vector
$\langle S_y{=}N | | S_x{=}N \rangle$. Another such example is the
measurement of the kinetic energy of a tunneling particle.$^8$ We have shown
for any precision of the measurement that we can ensure a negative value
reading of the measuring device by an appropriate choice of the
post-selection state.

However, these examples do not represent a measurement of the two-state
vector itself. If our measuring device for the spin measurement shows $\sqrt 2 N$,
we cannot deduce that our two-state vector is $\langle S_y{=}N | | S_x{=}N
\rangle$. Indeed, there are many other two-state vectors that yield the
same weak value for the spin component, but we cannot even claim that we
have one of these vectors. The probability for the result of the
post-selection measurement corresponding to any of these vectors is
extremely small, so it is much more likely to obtain the ``forbidden''
outcome $S_\xi=\sqrt 2 N$ as  a statistical error of the measuring
device. 
The same applies to the measurement of kinetic energy of a tunneling
particle. The negative value shown by the measuring device usually is due
to a statistical error, and only in very rare cases does it correspond to a
particle ``caught'' in the tunneling process.

We could try to use several weak measurements on a single pre- and
post-selected system in order to specify the two-state vector. But in
that case these measurements will change the two-state
vector. Therefore, as in the case of the measurement of the forward
evolving single-state vector of a single system, we need a protection
procedure.  At first sight, it seems that protection of a two-state
vector is impossible. Indeed, if we add a potential that makes one
state  a nondegenerate eigenstate, then the other state, if it is
different, cannot be an eigenstate too. (The states of the two-state
vector cannot be orthogonal.)  But, nevertheless, protection of the
two-state vector is possible, as we now  show.

 The procedure for protection of a two-state vector of a given system is
accomplished by coupling the system to another  pre- and post-selected
system. The protection procedure takes advantage of the fact that 
weak values  might  acquire complex values. Thus, the effective
Hamiltonian of the protection might not be Hermitian. Non-Hermitian
Hamiltonians act  in different ways on quantum states evolving forward and
backwards in time. This allows simultaneous protection of two different
states (evolving in opposite time directions).

Let us start with an example.$^9$ We consider  the protection of a  two-state
vector of a spin-1/2 particle, $\langle{\uparrow_y}
|  |{\uparrow_x} \rangle $. The protection procedure uses an external   pre- and post-selected
 system $S$ of a large  spin $N$ that is coupled to our spin via the interaction:
$$
H_{prot} = - \lambda {\bf S \cdot \sigma}. \eqno(3) 
$$
The external system is pre-selected in the state $|S_x {=} N\rangle$ and
post-selected in the state $\langle S_y {=} N|$, that is, it is described by
the two-state
vector $\langle S_y {=} N
|  |S_x {=} N \rangle $. The coupling constant $\lambda$ is chosen in such a
way that the
interaction with our spin-1/2 particle  cannot
change significantly the two-state vector of the protective system $S$, and
the spin-1/2 particle ``feels'' the effective Hamiltonian in which {\bf $S$} is
replaced by its weak value,
$$
{\bf S}_w = {{\langle S_y = N
|(S_x, S_y, S_z)  |S_x = N  \rangle} \over{\langle S_y = N
 |S_x = N  \rangle}} = (N, N, iN)  .\eqno(4)
$$  
Thus, the effective protective Hamiltonian is:
$$
H_{eff} =- \lambda N( \sigma_x +  \sigma_y + i\sigma_z). \eqno(5) 
$$
The state $|{\uparrow_x} \rangle $ is an eigenstates of this
(non-Hermitian) Hamiltonian (with eigenvalue $-\lambda N$).  For
backward evolving states the effective Hamiltonian is the hermitian
conjugate of (5) and it has different (nondegenerate) eigenstate with
this eigenvalue; the eigenstate is $\langle {\uparrow_y}|$.  The forward
evolving state $|{\uparrow_x} \rangle $ and the backward evolving
state $\langle {\uparrow_y} \vert $ are also the eigenstates of the
exact Hamiltonian (3) (when the large spin is pre- and post-selected
as described above).

 In order to prove that the Hamiltonian (3) indeed provides the
protection, we have to show that the two-state vector $\langle{\uparrow_y}
|   |{\uparrow_x} \rangle $ will remain essentially unchanged during the
measurement.
We consider measurement which is performed during the period of time,
between pre- and post-selection which we choose to be equal one. 
The Hamiltonian
$$
H = -\lambda {\bf S \cdot \sigma} + P  \sigma_{\xi} . \eqno(6) 
$$
can be replaced by the effective Hamiltonian: 
$$
H_{eff} =- \lambda N( \sigma_x +  \sigma_y + i\sigma_z) + P  \sigma_{\xi} . \eqno(7) 
$$
Indeed, the system with the spin $S$ can be considered as $N$ spin 1/2
particles all pre-selected in $|{\uparrow_x} \rangle$ state and
post-selected in $|{\uparrow_y} \rangle$ state. The strength of the
coupling to each spin 1/2 particle is $\lambda \ll 1$, therefore
during the time of the measurement their states cannot be changed
significantly. Thus, the forward evolving state 
$|S_x {=} N\rangle$ and  the  backward evolving state $\langle S_y {=}
N|$ do not change significantly during the measuring process. The
effective coupling to such system is the coupling to its weak values.

Good precision of the measurement of the spin component requires large
uncertainty in $P$, but we can arrange the experiment in such a way
that $P \ll N$. Then the second term in the Hamiltonian (6) will not
change significantly the eigenvectors. The two-state vector
$\langle{\uparrow_y} | |{\uparrow_x} \rangle $ will remain essentially
unchanged during the measurement, and therefore the measuring device
on this single particle will yield $({\sigma_\xi})_w =
{{\langle{\uparrow_y} |\sigma_\xi |{\uparrow_x} \rangle}\over
{\langle{\uparrow_y} |{\uparrow_x} \rangle}}$.  We can perform several
measurements of different spin component on the same single system
 since the measurements do not disturb significantly the
two-state vector.  Thus, the results $({\sigma_x})_w = 1$,
$({\sigma_y})_w = 1$, and $({\sigma_z})_w = i$ will uniquely define
the two-state vector.

 The Hamiltonian (3), with an  external system described
by the two-state vector $\langle S_y = N
|  |S_x = N \rangle $, provides protection for the two-state vector
$\langle{\uparrow_y}|  |{\uparrow_x} \rangle $. It is not difficult to
demonstrate 
that any two-state vector obtained by pre- and post-selection of  the
spin-1/2 particle can be protected by the Hamiltonian (3). A general form of
the two-state vector is $\langle{\uparrow_\beta}|  |{\uparrow_\alpha}
\rangle $ where  $\hat{\alpha}$ and $\hat{\beta}$ denote some directions. It can be
verified by a straightforward calculation that the two-state vector $\langle{\uparrow_\beta}|  |{\uparrow_\alpha}
\rangle $ is protected when  the two-state vector of the protective device
is  $\langle S_\beta = N
|  |S_\alpha = N \rangle $.

At least formally we can generalize this method to make a protective
measurement of an arbitrary two-state vector $\langle \Psi _2|  | \Psi
_1 \rangle $ of an arbitrary system.
 Let us decompose the post-selected state
 $| \Psi _2 \rangle = a | \Psi
_1 \rangle + b | \Psi _\bot \rangle $.
 Now we can define ``model spin'' states:
$ | \Psi _1 \rangle \equiv | \tilde{\uparrow}_z \rangle$ and $ | \Psi _\bot \rangle
\equiv | \tilde{\downarrow}_z \rangle$.
 On the basis of the two orthogonal states we
can obtain all other ``model spin'' states. For example,
$ |\tilde{\uparrow}_x \rangle = 1/\sqrt 2 ~( |\tilde{\uparrow}_z \rangle +
|\tilde{\downarrow}_z \rangle)$,
and then we can define  the ``spin model'' operator $\bf \tilde{\sigma}$. Now, the
protection Hamiltonian, in complete analogy with the spin-1/2 particle case
is 
$$
H_{prot} = - \lambda {\bf S \cdot \tilde{\sigma}}. \eqno(9) 
$$
In order to protect the state $\langle \Psi _2|  | \Psi _1 \rangle $, the
pre-selected state of the external system has to be $|S_z {=} N \rangle$ and the
post-selected state has to be  $\langle S_\chi {=} N|$ where the direction
$\hat{\chi}$ is defined by the  ``spin model'' representation of the state 
$| \Psi _2\rangle$:
$$|\tilde{\uparrow}_\chi \rangle \equiv | \Psi _2\rangle = \langle \Psi _1
| \Psi _2 \rangle|\tilde{\uparrow}_z \rangle + \langle \Psi _\bot
| \Psi _2 \rangle|\tilde{\downarrow}_z \rangle . \eqno(10)
$$

Let us come back to our first example. The Hamiltonian (5), has more
interesting features than just protecting the two state vector 
$\langle{\uparrow_y} | |{\uparrow_x} \rangle$. First, there is another
two-state vector which is protected: the two state
$\langle{\downarrow_x} | |{\downarrow_y} \rangle$ with corresponding
eigenvalue $\lambda N$. There is, however, a certain difference: while
$\langle{\uparrow_y} |$ and $|{\uparrow_x} \rangle$ are exact
eigenstates also of the  Hamiltonian (3) (with the chosen pre-
and post-selection of the spin $S$), the states $\langle{\downarrow_x}
|$, $ |{\downarrow_y} \rangle$ are not.  An
easy calculation shows that the
probability to find    $\sigma_y = 1$ at an intermediate time, given the initial state $
|{\downarrow_y} \rangle$, does not vanish, but it is small: the
probability is  of  order  $1/N^2$. 
 Straightforward (but lengthy) calculations show that the (not too strong) measurement coupling,  $P  \sigma_{\xi}$, adds to the
probability of finding  $\sigma_y = 1$ corrections proportional to
$P^2/N^2$, $P^2/\lambda^2N^2$, and   $P^4/\lambda^2N^2$ which are  
 also small for large  $N$. 

The calculations  show that $\lambda$ needs not be small for the
protection measurement. In fact,  larger $\lambda$ yields better
protection.   We required small $\lambda$ to ensure that the coupling
(3) will not cause significant change of the two-state of the large
spin $S$ system, irrespectively of the evolution of the spin-1/2
particle. But, when the additional coupling $ P  \sigma_{\xi}$ is
small compare to the protection Hamiltonian (3), the spin-1/2 particle
evolves in such a way that the two-state vector $\langle S_y=N | | S_x =
N \rangle$ remains essentially unchanged even when $\lambda$ is large. 

  Another important point is that the bound on $P$, and thus the bound
on the precision of the measurement, can be reduced by increasing the
period of time $T$ of the measurement with the appropriate reduction of
the strength of the coupling term, $ P  \sigma_{\xi}/T$. For this
regime we can give another proof that our intermediate measurements
yield the weak values.$^{10}$

In general, a nondegenerate nonhermitian Hamiltonian can be written in
the following form
$$
H= \Sigma_i \omega_i {{|\Phi_i\rangle \langle \Psi_i |}\over {\langle
\Psi_i| \Phi_i\rangle}} ,\eqno(11)
$$
where $\langle \Psi_i|$ are the ``eigen-bras'' of $H$, and
$|\Phi_i\rangle$ are the ``eigen-kets'' of $H$. The $\langle
\Psi_i|$ form a complete but, in general, non-orthogonal set, and so do
the $|\Phi_i\rangle$. They obey mutual orthogonality condition:
$\langle \Psi_i|\Phi_i\rangle = \delta_{ij} \langle
\Psi_i|\Phi_i\rangle$.

The Hamiltonian of our example  gets the form
$$
H_{eff} = - \lambda N {{|\uparrow_x\rangle \langle \uparrow_y|}
\over {\langle\uparrow_y|\uparrow_x\rangle} }
+  \lambda N {{|\downarrow_y\rangle \langle \downarrow_x|}
\over {\langle\downarrow_x|\downarrow_y\rangle} }
+{P\over T} \sigma_\xi  .\eqno(12)
 $$
Diagonalisation of the Hamiltonian yields the modified energy
eigenstates
$$
\omega_1 = - \lambda N + {P\over T}
{{\langle{\uparrow_y} |\sigma_\xi |{\uparrow_x} \rangle}\over
{\langle{\uparrow_y} |{\uparrow_x} \rangle}}, ~~ \omega_2 =  \lambda N + {P\over T}
{{\langle{\downarrow_x} |\sigma_\xi |{\downarrow_y} \rangle}\over
{\langle{\downarrow_y} |{\downarrow_x} \rangle}}. \eqno(13)
$$
This means   that if the initial state of the system is
$|\uparrow_x\rangle$, then the measuring device will record the weak value
of $\sigma_\xi$ for the two-state vector $\langle \uparrow_y |
|\uparrow_x\rangle$. This result even stronger  than what we wanted to
show since we do not
require the
post-selection of  the state  $\langle \uparrow_y|$. The reason why
other components of the backward evolving state do not contribute is
because the corresponding component of the forward evolving state has
zero amplitude. This feature will be clearer after the following
discussion.

It is interesting to analyze the behavior of a system described by
nonhermitian Hamiltonian (11) when the initial state is not one of the
eigenstates. In this case the initial state should be decomposed into
a superposition of the eigenstates $|\Psi\rangle = \Sigma_i \alpha_i |\Psi_i\rangle$
and its time evolution will be given by 
$$
|\Psi (t)\rangle ={\cal N}(t) \Sigma_i \alpha_i e^{-i\omega_i T}| \Psi_i\rangle
\eqno(14)
$$
In order to keep the state normalized we have to introduce the time
dependent normalization factor ${\cal N}(t)$.  This is the difference
in the action of the the effective
Hamiltonian, and it  signifies the fact that the probability
for the appropriate result of the post-selection measurement (which
leads to the nonhermitian effective Hamiltonian) depends on the time
when it is performed. 

If an adiabatic measurement of a variable $A$ is performed then the
final state of the system and the measuring device is  
$$
{\cal N'}(t) \Sigma_i \alpha_i e^{-i\omega_i T}| \Psi_i\rangle \Phi (Q-{{\langle \Phi_i | A | \Psi_i \rangle}
\over {\langle \Phi_i |\Psi_i \rangle}}) .\eqno(15)
$$
The state of the measuring device is amplified to a macroscopically
distinguishable situation and, according to standard  interpretation,  a
collapse takes place to  the reading of one of the {\it weak values}
of $A$ with the relative probabilities given by $|\alpha_i
e^{-i\omega_i T}|^2$.

 In general, construction of the formal protection Hamiltonian (9)
which leads to the nonhermitian Hamiltonian is a gedanken experiment.
It generates nonlocal interactions which can contradict relativistic
causality. However, effective nonhermitian Hamiltonian can be obtained
in a real laboratory in a natural way when we consider  a decaying system
and we post-select the cases in which it did not decayed during the
period of time $T$ which is  larger than its
characteristic decay time. Kaon decay is  such an example. $|K_L^0\rangle$ and
$|K_S^0\rangle$ are the eigen-kets of the effective Hamiltonian and they have
corresponding eigen-bras  $\langle K_L'^0|$ and
$\langle {K'}_S^0|$ evolving backward in time. Due to the
$CP- violation$ the states  $|{K}_L^0\rangle$ and
$|K_S^0\rangle$   are not orthogonal. However, the mixing is
small:
$|\langle K_S^0|K_L^0\rangle|\ll 1$, and therefore the  
corresponding backward evolving states are almost identical to the
forward evolving states: $|\langle {K'}_S^0|K_S^0\rangle| = |\langle
{K'}_L^0|K_L^0\rangle| = {1\over
{\sqrt {1 -|\langle
K_S^0|K_L^0\rangle|^2}}}$. Thus, it is difficult to expect a large
effect in this system and for a realistic experimental proposal one
should look, probably, for another system.

We have shown in the framework of nonrelativistic quantum theory that we
can measure (or, maybe a better word, ``observe'') two-state vectors
describing pre- and post-selected 
quantum systems. A number of such 
measurements define the two-state vector and we have a procedure to
protect the two-state vector from significant change due to these measurements.
In order to protect, we have to know the two-state vector. Thus, this
procedure is liable to a criticism$^{11-13}$ leveled at  our first
proposal. Our response  to this can be found in Ref. 14.  Although we
consider our present proposal to be a measurement
performed on a single system, it should also be
mentioned  that in any realistic practical implementation
 we will need  ensembles of particles, protective systems, and  measuring
devices.  The external system of the protective
device has to be not only prepared (pre-selected) in a certain state, but
also post-selected in a given state. In all interesting cases  the
probability for an appropriate outcome of the post-selection measurement is
extremely small. Still, there is a non-zero probability that our first run
with a single system, a single protective device, and a single set of measuring
devices will yield the desired outcomes. In this case we have a reliable measurement
performed on a single system. However, even when  we  use  a pre-selected
ensemble,  we actually use only  a single pre- and post-selected
system. After achieving the first successful post-selection, we have
completed  the
experiment. For more discussion of this point, see Ref. 15.

It is interesting to notice that our procedure cannot protect a {\it
generalized two-state vector}$^7$ which is a superposition of
two-state vectors. The system described by a generalized two-state
vector is correlated to some external system. It seems that it is
impossible to find any protective procedure of the generalized
two-state vector that does not involve coupling to that external
system. This feature hints that the generalized two-state vector,
although useful as a tool, is not a basic concept. The composite
system consisting of the system under study and the system correlated
to it is described by the usual, basic two-state vector.

We have shown that the two-state vector is observable. Previously we have
shown that the single-state vector is observable. For weak coupling
interactions to an observable $A$, in the first approach we obtain the effective coupling  to the weak
value 
  ${{\langle \Psi_2 | A | \Psi_1 \rangle}
\over {\langle \Psi_2 |\Psi_1 \rangle}}$, while in the second,  to
the expectation value $\langle \Psi_1 | A | \Psi_1 \rangle$. Since the two values
are, in general different, we encounter an apparent paradox. The
resolution of the paradox is as follows:

In order to observe a quantum state it has to be protected. When we
discussed the protective experiments of single-state vectors we did
not say anything about quantum states evolving backwards in time. (It
was not related to the point we wanted to make.) However, the
protective procedure that we proposed automatically protects the {\it
identical} state evolving backward. Thus, what we have proposed as an
observation of a single-state vector is in fact an observation a
two-state vector with identical forward and backward evolving
states. For example, the protection of spin-1/2 particle state,$^2$ a
strong magnetic field in a given direction, protects the two-state
vector with either both states parallel or anti-parallel to this
direction. This procedure is incompatible with the protection of the
forward evolving state parallel to one direction and the backward
evolving state parallel to another. If the particle is described by
$\langle{\uparrow_y} | |{\uparrow_x} \rangle $ then the strong
magnetic field in the $\hat{x}$ direction will change the backward
evolving spin-state. There exists a protection procedure for $
|{\uparrow_x} \rangle $ that does not change the backward evolving
state as was described in the preceding section. The ``observation''
of the state protected in such a way will not yield the pre-selected
quantum state but it will yield the picture defined by the two-state
vector.

  The research was supported in part by grant 614/95 of
 the Basic Research Foundation (administered by the Israel Academy
of Sciences and Humanities).

\vskip 0.7 cm

\centerline{\bf
 REFERENCES}
\vskip .2  cm
 
\noindent
1. Y. Aharonov and L. Vaidman, {\it Phys. Lett.} {\bf A178}, 38
(1993).\hfill \break
2. Y. Aharonov, J. Anandan, and L. Vaidman, {\it Phys. Rev.} {\bf
 A 47}, 4616 (1993).\hfill \break
3. Y. Aharonov, P.G. Bergmann  and J.L.
   Lebowitz,   {\it  Phys.  Rev.}  {\bf B134},  1410 (1964).\hfill \break
4. Y. Aharonov, D. Albert, A. Casher, and L. Vaidman,
 {\it Phys. Lett.} {\bf  A 124}, 199 (1987).\hfil \break
5.  Y.Aharonov and L. Vaidman, {\it Phys.  Rev.}   {\bf A 41}, 11 (1990).
\hfil \break
6.  Y.Aharonov and L. Vaidman, {\it J. Phys.} {\bf A 24}, 2315
(1991).\hfil \break
7.  B. Reznik and Y.Aharonov  {\it Phys.  Rev.}   {\bf A
    52}, 2538 (1995).
\hfil \break
8. Y. Aharonov,  S. Popescu, D.~Rohrlich, and L.~Vaidman, {\it Phys.  Rev.}
{\bf A 48}, 4084 (1993).\hfill \break
9.  Y.Aharonov and L. Vaidman, {\it Ann. NY Acad. Sci.}
  {\bf 755}, 361 (1995).\hfill \break
10.  Y.Aharonov, S. Massar, S. Popescu, J. Tollaksen, and
  L. Vaidman, TAUP 2315-96.\hfill \break
11. W. G. Unruh, {\it Phys. Rev.} {\bf A 50}, 882 (1994). \hfill  \break
12. C. Rovelli, {\it Phys. Rev.} {\bf A 50}, 2788 (1994). \hfill  \break
13. P. Ghose and D. Home, {\it  Found. Phys.} {\bf 25}, 1105 (1995).\hfill \break
14. Y. Aharonov, J. Anandan, and L. Vaidman, ``The Meaning of the Protective
Measurements,'' {\it Found. Phys.} to be published.\hfill \break
15. L. Vaidman, 
in {\it Advances in Quantum Phenomena}, E. Beltrametti and J.M.
Levy-Leblond eds.,  NATO ASI series, Plenum, NY (1995). 
\end